\documentclass[conference]{IEEEtran}

\usepackage[cmex10]{amsmath}
\usepackage{mathtools}
\usepackage{amsfonts}
\usepackage{amssymb}
\usepackage{cite}

\usepackage{breqn}

\usepackage[pdftex]{graphicx}
\graphicspath{{images/}}
\usepackage[absolute,showboxes]{textpos}

\usepackage{tikz}
\usetikzlibrary{arrows,matrix}

\usepackage[absolute,showboxes]{textpos}

\TPMargin{5pt}

\def\eps{\varepsilon}

\usepackage{flushend}
\usepackage{caption}
\usepackage{subcaption}

\usepackage{multicol}
\usepackage{color}

\DeclareMathOperator{\Prob}{\mathbb{P}}
\DeclareMathOperator{\E}{\mathbb{E}}

\usepackage[absolute,showboxes]{textpos}

\TPMargin{5pt}

\IEEEoverridecommandlockouts

\begin{document}

\title{On the Reliability of LTE Random Access: Performance Bounds for Machine-to-Machine Burst Resolution Time\vspace{-.4cm}}

\author{\IEEEauthorblockN{Mikhail Vilgelm\IEEEauthorrefmark{1}, Sebastian Schiessl\IEEEauthorrefmark{2}, Hussein Al-Zubaidy\IEEEauthorrefmark{2}, Wolfgang Kellerer\IEEEauthorrefmark{1}, James Gross\IEEEauthorrefmark{2}\\
\IEEEauthorrefmark{1} Chair of Communication Networks, Technical University of Munich, Germany \\\IEEEauthorrefmark{2} School of Electrical Engineering, KTH Royal Institute of Technology, Stockholm, Sweden\\
Email: \IEEEauthorrefmark{1}\{mikhail.vilgelm, wolfgang.kellerer\}@tum.de, \IEEEauthorrefmark{2}\{schiessl, hzubaidy, jamesgr\}@kth.se}\vspace{.1cm}\thanks{This work was in part supported by the German Research Foundation (DFG) grant KE1863/5-1 as part of the priority program SPP1914 Cyber-Physical Networking.}}

\maketitle
\begin{abstract}
Random Access Channel (RACH) has been identified as one of the major bottlenecks for accommodating massive number of Machine-to-Machine (M2M) devices in LTE networks, especially in the case of bursty arrivals of connection requests. As a consequence, the burst resolution problem has sparked a large number of works analyzing and optimizing the expected performance of RACH. In this paper, we go beyond the study of performance in expectation by investigating the probabilistic performance limits of RACH with access class barring. We model RACH as a queuing system, and apply stochastic network calculus to derive probabilistic performance bounds for burst resolution time, i.e., the time it takes to connect a burst of M2M devices to the base station. We illustrate the accuracy of the proposed methodology and its potential applications in performance assessment and system dimensioning.
\end{abstract}
\IEEEpeerreviewmaketitle

\section{Introduction}

One of the main goals for the evolution from current wireless systems to 5G is to support massive Machine-to-Machine (M2M) communications. This could enable a number of promising applications, such as large scale sensor and actuator networks, smart grid monitoring, and many more~\cite{6815890}. Main challenges for enabling massive M2M are: improving scalability, enhancing energy efficiency and decreasing the cost of user equipment (UE). In this work, we focus on a notorious scalability issue of nowadays LTE networks, namely, on the connection establishment procedure. For many M2M scenarios, connection establishment dominates the end-to-end delays, hence, creating a bottleneck in the Random Access CHannel (RACH) even before the actual data transmission begins~\cite{laya2014random}. The issue becomes especially critical in the case of the simultaneous activation of a large group of UEs, e.g., sensors re-connecting after a power outage~\cite{TR37868}. This simultaneous triggering is referred to as a \textit{burst arrival}, and it creates a RACH overload persisting over a long time.

Burst arrivals has been a known problem for random access protocols since the early works in the field~\cite{firstBatch1988}. Unlike independent uncorrelated arrivals, bursts could significantly degrade the performance of random access protocols. Apart from LTE for M2M, bursty behaviors were studied in the context of sensor and RFID networks~\cite{popovski2004batch}. Many algorithms to ``smoothen'' the negative effect of the bursts have been developed~\cite{laya2014random}, for instance, by means of back-off or barring, e.g., pseudo-Bayesian broadcast~\cite{rivest1987network}, or tree resolution algorithms~\cite{popovski2004batch}. For LTE, standardized mechanism to deal with burst arrivals is Access Class Barring (ACB)~\cite{TR37868}: re-shaping the burst by broadcasting a RACH access probability for all UEs.

In the state-of-the-art, the performance of ACB and its derivatives~\cite{duan2016d,7875393} has been extensively studied \textit{in the expectation}, i.e., with respect to the \textit{average} burst resolution time and resulting RACH efficiency~\cite{wei2015modeling,cheng2015modeling,7875393,7447749,7577764}. In~\cite{wei2015modeling} and the follow-up work~\cite{cheng2015modeling}, the authors devised an analytical framework to assess the expected performance of the standardized ACB and Extended Access Class Barring (EAB) procedures, respectively. Jian~\textit{et al.}~\cite{7577764} have proposed another iterative approach to the ACB analysis, and Koseoglu~\cite{7447749} derived the lower bound on the average random access delay. 

However, for many applications on the border between massive M2M and ultra reliable M2M~\cite{popovski2014ultra}, e.g., in-cabin communication in an aircraft or large-scale industrial automation~\cite{7870628}, assessing the average performance is insufficient. For instance, if all the sensors in a factory need to re-connect after an emergency shutdown within a certain time limit~\cite{7870628}. In that case, \textit{reliability guarantees for the RACH performance} are necessary. As a first step towards designing the reliable random access procedures for such scenarios, our work aims to answer the question of what the performance limits of the existing standardized solutions are.

In this paper, we analytically study the probabilistic performance bounds of standardized LTE RACH with ACB. We investigate the \textit{burst resolution time}, i.e., the time it takes to connect a burst of M2M devices to the base station. Modeling RACH as a queuing system, we approach the analysis by the means of stochastic network calculus~\cite{fidler2006end}, which allows, in contrast to conventional queuing theory, to characterize the behavior of the system in probability and not only in expectation~\cite{ciucu2014towards}. We analyze what burst resolution delay can be guaranteed for a given burst size with a certain reliability requirement. We validate the approach using simulations, and illustrate possible applications of the proposed methodology.

The remainder of the paper is structured as follows. We introduce the problem and relevant concepts in~\ref{sec:system}. The main result of the paper, probabilistic reliability analysis of the random access procedure is presented in~\ref{sec:queuinganalysis}, and numerically verified in~\ref{sec:evaluation}. We conclude the paper with~\ref{sec:conclusions}.

\section{System Model and Preliminaries}
\label{sec:system}

\subsection{System Model}

We consider a scenario with a total of $N$ UEs and one base station (BS). At time $t<0$, all UEs are inactive and disconnected from the BS. An event is occurring at time $t=0$, triggering all the UEs, and causing them to initiate a connection establishment (random access) procedure towards the BS. Activation of individual UEs is occurring according to initial arrival process $A(t)$ strictly during the time interval $t\in[0,T_A-1]$, with $T_A$ referred to as the activation time~\cite{TR37868}.

\begin{figure}[t!]
	\centering
	\includegraphics[width=1\linewidth]{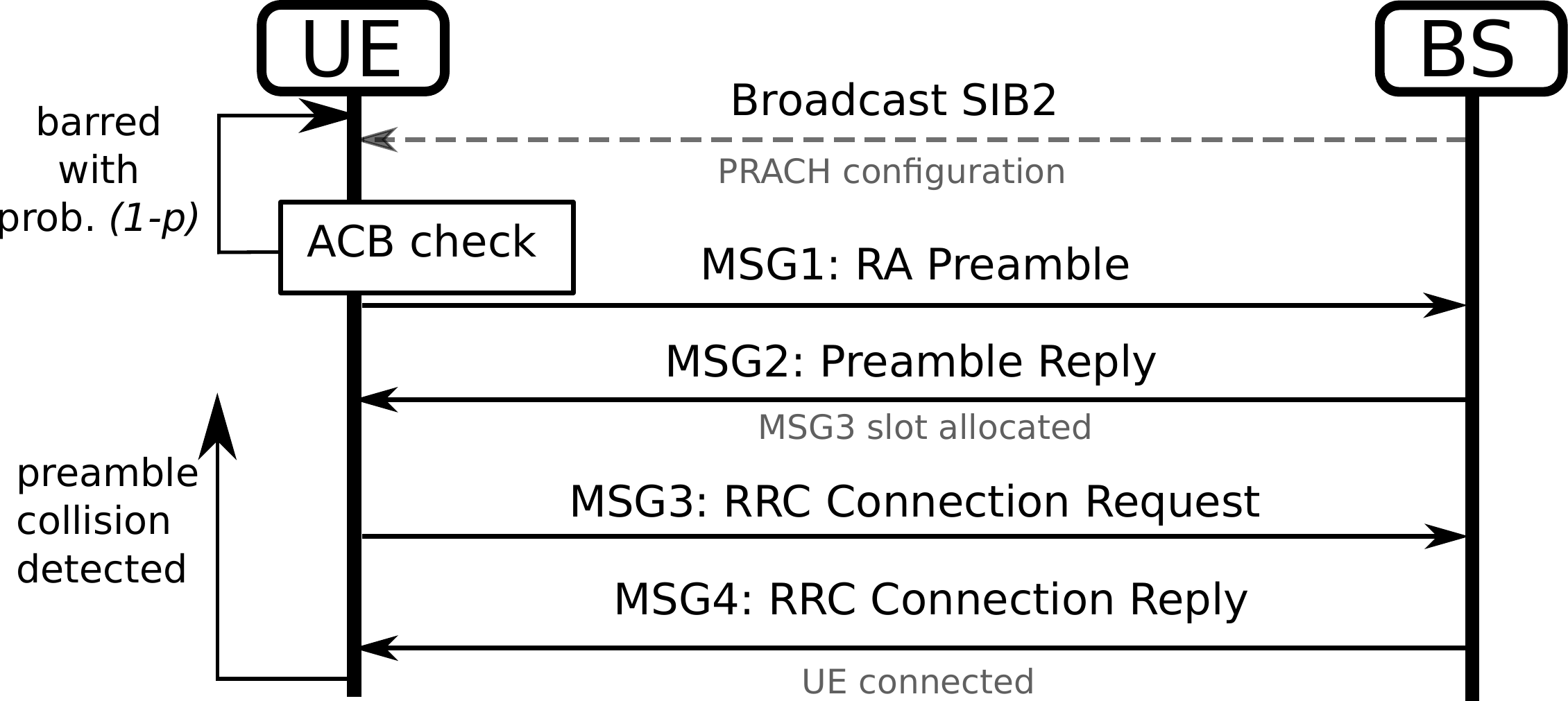}
	\caption{Four steps LTE Random Access procedure with ACB.}
	\label{fig:rachprotocol}
	\vspace{-0.6cm}
\end{figure}

Upon activation, every UE attempts to connect to the BS. The connection follows a four step random access procedure, depicted in Fig.~\ref{fig:rachprotocol}: (1) A preamble, chosen uniformly random from a set $|\mathcal{M}|=M$, is sent to the BS in Physical Random Access Channel (PRACH). (2) The BS sends a preamble reply for every successfully decoded preamble, containing uplink grants for RRC connection requests. (3) UE proceeds with sending its connection request, containing UE identity information, on the respective uplink resource. (4) Every correctly decoded connection request is acknowledged by the BS with a connection reply. If the UEs choose the same preamble in step 1, their connection requests at step 3 are allocated the same uplink resource, which leads to collisions.

Prior to every PRACH attempt $i$, UEs receive the PRACH location (sub-frame and frequency offset), and contention parameters (number of available preambles $M$, access probability $p_i$) from a BS broadcast. Every UE independently uses access probability as a part of the ACB procedure to decide whether to compete in a given PRACH opportunity $i$ (with probability $p_i$), or postpone to the next one (with probability $1-p_i$). The access probability could be either static throughout the burst resolution, or dynamically adapted for every PRACH opportunity~\cite{duan2016d}.

We assume that all four steps occur within one \textit{PRACH slot}, which we define as periodicity of the PRACH in the LTE resource grid. The periodicity is determined by the PRACH configuration index, typically ranging from 1 per sub-frame ($1$~ms) to 1 per frame ($10$~ms). Random access procedure is modeled as an \textit{$M$-channel slotted ALOHA protocol}, where a channel corresponds to a PRACH preamble~\cite{wei2015modeling}. We further adopt the \textit{collision channel model without capture}, i.e., every preamble $m$ at time slot $i$ can have one of three states: \textit{idle} (no UE is choosing the preamble), \textit{singleton} (exactly $1$ UE), and \textit{collision} ($\geq 2$ UEs), with a corresponding service $s_{i,m}$:
\begin{equation}
s_{i,m} = \begin{cases}
1 & \text{if chosen by 1 UE,}\\
0 & \text{otherwise.}
\end{cases}
\label{eqn:service_per_preamble}
\end{equation}
Basically, a preamble is serving a UE request with a full channel capacity if no collision occurs, and does not provide any service otherwise. Every activated, but not yet served UE is denoted as \textit{backlogged}. 

\subsection{Problem Statement}

Finally, we define the target quality of service (QoS) requirement\footnote{Typically, QoS is defined per single user/application, and refers to the delay or datarate requirement. In contrast to that, we are analyzing the burst resolution (multiple users), and borrow the term QoS requirement to refer to the backlog and resolution time.} of the system as a tuple $(b^\eps, t, \eps)$. Here, $b^\eps$ is the maximum tolerated number of unconnected UEs (\textit{target backlog}) by the time $t$, which is referred to as the required burst resolution time. The burst is denoted as \textit{resolved} if the number of unconnected UEs is less than or equal than target, $B(t)\leq b^\eps$. The corresponding unreliability $\eps$ is the probability that a burst is not resolved by the time $t$. The case with $b^\eps=0$ corresponds to the full burst resolution, and $b^\eps>0$ to the partial burst resolution~\cite{popovski2004batch}.

The problem we are targeting with the analysis is quantifying how well can the ACB-based random access procedure support a given QoS requirement, i.e., for a given target $b^\eps$, we would like to compute a bound on the probability $\eps$ that a given burst is not resolved within $t$ PRACH slots.

\subsection{Analysis Preliminaries}
\label{sec:prelims}

Consider the system at an arbitrary time slot $i\geq 0$ with $B(i)$ backlogged UEs. The evolution of the backlog is described by the following recursion:
\begin{equation}
B(i+1) = \max\{0,B(i) + a_i - s_i\},
\label{eqn:backlogevolution}
\end{equation}
where $a_i$ denotes newly activated UEs, and $s_i = \sum_{m=1}^{M}s_{i,m}$ denotes the total amount of served UEs.

The probability that $s_i=k$ UEs are served during the slot $i$, i.e., have successfully connected to the BS, depends on current backlog $B(i)$ and access probability $p_i$, limiting the number of admitted for contention UEs $b^\prime_i$. Let us first consider that the number of admitted UEs is $b^\prime_i=x$. In that case, the probability that $k$ out of $x$ UEs successfully transmit is~\cite{wei2015modeling,duan2016d}:
\begin{flalign}
\Prob&\left[s_i=k|b_i^\prime=x\right]=\binom{x}{k}\binom{M}{k}\frac{k!}{M^x}\text{ }\times\\
&\times\sum_{j=1}^{j_{\max}}(-1)^j\binom{M-k}{j}\binom{x-k}{j}j!(M-k-j)^{x-k-j},\nonumber
\end{flalign}

where $j_{\max}=\min(M-k,x-k)$. The number of admitted UEs $b_i^\prime$ is binomially distributed with $B(i)$ trials and per trial success probability $p_i$:
\begin{equation}
\Prob\left[b_i^\prime=x|B(i)=n\right] =\binom{n}{x}(1-p_i)^xp_i^{n-x}.
\end{equation}

Combining these two equations, we obtain the probability $\Prob_{k,n}=\Prob[s_i=k|B(i)=n]$ that $k$ out of $n$ backlogged UEs are successful as:
\begin{equation}
\Prob_{k,n}= \sum_{x=0}^{n}\Prob\left[b_i^\prime=x|B(i)=n\right]\Prob\left[s_i=k|b_i^\prime=x\right].
\label{eqn:transition_prob_sending}
\end{equation}

Eqn.~\eqref{eqn:transition_prob_sending} already allows a straightforward recursive computation of the burst resolution time. If we consider a state of the system at time $i$ as a tuple $(B(i),a_i)$, where $B(i),a_i\in[0,N]$, then the distribution of the random variable $B(i)$ representing backlog at time $i$ can be computed iteratively starting with $t=0$, using the recursion~\eqref{eqn:backlogevolution}. However, this iterative computation requires computing transition matrix from $(N+1)\times(N+1)$ to another $(N+1)\times(N+1)$ dimension state space every time step, and, hence, the complexity is proportional to $(N+1)^2\times(N+1)^2\times t$. Such computation is only feasible for low total number of UEs $N$.
For large bursts, a different approach is necessary. This motivates the network calculus based analysis, which we present in Sec.~\ref{sec:queuinganalysis}.

\subsection{Dynamic Access Barring}
For large burst arrivals, keeping access probability static is very inefficient. If the probability is too small, burst resolution lasts long due to the medium under-utilization as the backlog decreases. If the access probability is too large, the burst resolution might take even longer due to high preamble collision rates. To optimize the burst resolution times, several works have proposed a dynamic adaptation of the access probability~\cite{duan2016d,7875393}, based on the pseudo-Bayesian broadcast~\cite{rivest1987network}. Consider expected number of successful UEs in a single slot, as a function of $b_i^\prime,M$~\cite{wei2015modeling}:
\begin{equation}
\E\left[s_{i}\right] = \E\left[\sum_{m\in\mathcal{M}}s_{i,m}\right] = \E[b_i^\prime]\left(1-\frac{1}{M}\right)^{\E[b_i^\prime]-1}.
\label{eqn:exp_success}
\end{equation}
It is possible to show that the expectation $\E\left[s_{i}\right]$ in~\eqref{eqn:exp_success} is maximized if the expected number of devices admitted to contend in a given slot $\E[b_i^\prime]=p_iB(i)$ is equal to the number of preambles $M$. Hence, the dynamic access barring policy is devised as:
\begin{eqnarray}
p_i^\star\triangleq\arg\max_{p\in(0,1]}\E\left[\sum_{m\in\mathcal{M}}s_{i,m}\right]
=\min\left\lbrace 1,\frac{M}{B(i)}\right\rbrace.\label{eqn:optimalacb}
\end{eqnarray}
We denote $p_i^\star$ defined by~\eqref{eqn:optimalacb} as \textit{optimal barring policy}. 

\textbf{Remark.} In general, the number of UEs contending in a given PRACH slot $B(i)$ is unknown. However, there exist a number of backlog estimation techniques, producing accurate results~\cite{duan2016d,popovski2004batch,7875393,7078932,6134704}. We study the impact of estimation numerically in Sec.~\ref{sec:evaluation}.
\section{Probabilistic Burst Resolution Time Analysis}
\label{sec:queuinganalysis}

In this section, we present the burst resolution time analysis. To introduce the reader to stochastic network calculus, we first provide a brief overview in~\ref{Sec:Basic_Model}. Then, we define the queuing model of LTE RACH in~\ref{sec:rachqueuing}, and use it to analyse static (in~\ref{subsec:static}) and dynamic (in~\ref{subsec:dynamic},~\ref{sec:fullburst}) ACB policies.

\subsection{Transient Analysis using Network Calculus}
\label{Sec:Basic_Model}
Assuming a fluid-flow, discrete-time queuing system, and given a time interval $[s,t)$, $0\le s \leq t $, we define the non-decreasing (in $t$) bivariate processes $A(s,t), D(s,t)$ and $S(s,t)$ as the cumulative arrival to, departure from and service offered by the system. 
We further assume that $A,D$ and $S$ are stationary non-negative random processes with $A(t,t)= D(t,t)= S(t,t)=0$ for all $t\ge 0$.
The cumulative arrival and service  processes  are given in terms of $ a_{i} $ and $ s_{i} $ as follows
\begin{equation}\label{eq:cumulative-arrival-service}
A(s,t)=\sum_{i=s}^{t-1}a_{i}\; \mbox{ and } \; S(s,t)=\sum_{i=s}^{t-1}s_{i} \, ,
\end{equation}
for all $0 \le s \le t$. 
We denote by $B(t)$ the backlog (the amount of buffered data) at time $t$.


Based on this server model, the total backlog can be studied analytically. 
For a given queuing system with cumulative arrival $ A(0,t) $ and departure $ D(0,t) $ and for $t\ge 0$, the backlog at time $t$, $ B(t) $ is defined as the amount of traffic remaining in the system by time $t$. Therefore,
\begin{equation}\label{eqn:backlog1}
B(t)\triangleq A(0,t)-D(0,t)\, .
\end{equation}
While deterministic network calculus~\cite{le2001network} can provide worst-case upper bounds on the backlog and the delay if traffic envelopes (an upper bound on the arrival process) as well as a service curve (a lower bound on the service process) are considered, probabilistic performance bounds provide more useful and realistic description of the system performance than deterministic analysis for corresponding systems. Stochastic network calculus has been previously applied to protocol analysis in the context of 802.11 DCF and slotted ALOHA networks~\cite{ciucu2014towards,poloczek2015service}.

%
%
%
%
In the probabilistic setting (where the arrival process $A$ and the service process $S$ are stationary random processes), the backlog defined in \eqref{eqn:backlog1} is reformulated in a stochastic sense:
\begin{equation}\label{eqn:probabilistic bounds}
\mathbb{P}[B(t)>b^{\eps}]\leq \eps,
\end{equation}
where $ b^{\eps} $ denotes the target probabilistic backlog associated with violation probability $\eps$.
This performance bound can be obtained by the distributions of the processes, i.e., in terms of moment generating functions (MGFs) of the arrival and service processes~\cite{fidler2010survey}.
In general, the MGF-based bounds are obtained by applying Chernoff's bound, that is, given a random variable $X$, we have  
\begin{align*}
\mathbb{P}[X \geq x]\leq e^{-\theta x} \mathbb{E}\left[e^{\theta X}\right]=e^{-\theta x}\mathbb{M}_{X}(\theta),
\end{align*}
whenever the expectation exists, where $ \mathbb{E}\left[Y \right] $ and $ \mathbb{M}_{Y}\left(\theta\right) $ denote the expectation  and the MGF (or the Laplace transform) of $ Y $, respectively, and  $ \theta $ is an arbitrary non-negative free parameter. Given the stochastic process $ X(s,t),$ $t\geq s$, we define the MGF of $ X $ for any $ \theta \geq 0 $ as~\cite{fidler2006end}
\begin{align*}
\mathbb{M}_{X}(\theta,s,t)\triangleq\mathbb{E}\left[e^{\theta X(s,t)}\right].
\end{align*}
In a similar way, we define $\overline{\mathbb{M}}_{X}(\theta,s,t)\triangleq \mathbb{M}_{X}(-\theta,s,t)= \mathbb{E}\left[e^{-\theta X(s,t)}\right] $.

A number of properties of MGF-based network calculus are summarized in \cite{fidler2006end}.  
In this work, we consider a queuing system with an initial backlog for which we are interested in the transient behavior of the backlog itself. 
In general, the probabilistic backlog bound $b^{\eps}(t)$ for a given violation probability $\eps$ at time $t$ can be expressed by \cite{fidler2006end, al2013min}
\begin{equation}
b^{\eps}(t)=\inf_{\theta >0}\left\lbrace\frac{1}{\theta}\left(\log \mathsf{M}(\theta,t,t)-\log \eps \right) \right\rbrace,
\label{eqn: backlog bound}
\end{equation}
where $ \mathsf{M}(\theta,u,v) $ is given as  
\begin{equation} 
\mathsf{M}(\theta,u,v)\triangleq\sum_{k=0}^{\min(u,v)}\mathbb{M}_{A}(\theta,k,v)\cdot \overline{\mathbb{M}}_{S}(\theta,k,u).
\label{eqn: generic MGF for backlog and delay}
\end{equation}
The consideration of the initial backlog in the system can be finally represented by the choice of an appropriate arrival function, as we discuss in the next section.

\subsection{Queuing Model of Random Access Procedure}
\label{sec:rachqueuing}
Random Access Procedure could be viewed as a queuing system, where the incoming UEs are considered as \textit{arrivals} into the queue, and UEs, successfully completing the procedure, as \textit{departures} from the queue. The serving process of such a system is a stochastic process, dependent on the current backlog size, on the advertised access probability $p_i$, and on the number of available preambles $M$, as in Eqn.~\eqref{eqn:transition_prob_sending}.

\subsubsection{Arrival Process}

3GPP offers three burst arrival models~\cite{TR37868}: delta (simultaneous, ``spike'') arrivals with total activation time $T_A=0$, uniform distribution of arrivals within $[0,T_A-1]$, or beta distribution $\mathsf{B}(\alpha,\beta)$ within $[0,T_A-1]$. In this works, we consider only the worst-case scenario with simultaneous activation of all UEs. In that case, the distribution of the activation time $t_a$ of individual UEs and the resulting cumulative arrival process are expressed as:
\begin{equation}
\Prob[t_a=0] = 1\quad\text{and}\quad A(\tau,t) = \begin{cases}
N & \tau=0,\\
0 & \tau>0.
\end{cases}
\end{equation}
%

\subsubsection{Serving Process}
Every preamble $m\in\mathcal{M}$ can be considered a server, with the service as defined in~\eqref{eqn:service_per_preamble}. Hence, cumulative serving process can be expressed as:
\begin{align}
S(\tau,t) &\triangleq \sum_{i=\tau}^{t-1}\sum_{m\in\mathcal{M}}s_{i,m},\\
\text{with }&s_{i}\triangleq\sum_{m\in\mathcal{M}}s_{i,m}\sim f_i(k),
\end{align}

where the probability mass function (PMF) of the service process at the time step $i$ as $f_i(\cdot)$:
\begin{align}
	f_i(k)&=\sum_{j=0}^{N}\Prob_{k,j}\Prob[B(i-1)=j]\nonumber\\
	&=\sum_{j=0}^{N}\Prob_{k,j}\sum_{l=0}^{N}\Prob_{l-j,l}\Prob[B(i-2)=l]\nonumber\\
 	&=\sum_{j=0}^{N}\Prob_{k,j}\sum_{l=0}^{N}\Prob_{l-j,l}\sum_{m=0}^{N}\Prob_{m-l,m}\Prob[B(i-3)=m]\nonumber\\
&=\sum_{j=0}^{N}\Prob_{k,j}\dots\sum_{y=0}^{N}\Prob_{y-x,y}\Prob[B(0)=y]\nonumber\\
	&=\underbrace{\sum_{j=0}^{N}\Prob_{k,j}\dots\sum_{y=0}^{N}\Prob_{y-x,y}}_{i\text{ sums}}\Prob_{N-y,N}
	\label{eqn:servicedist}
\end{align}

Similarly, we derive the distribution of the cumulative service $S(0,i)$, $f_{S_i}$ as
\begin{align}
&f_{S_i}(k)=\Prob[S(0,i)=k]=\label{eqn:cumservicedist}\\
&=\sum_{j=\max(0,k-M)}^{k}\Prob_{k-j,N-j}\Prob[S(0,i-1)=j]=\nonumber\\
&=\sum_{j=\max(0,k-M)}^{k}\Prob_{k-j,N-j}\times\nonumber\\
&\times\sum_{n=\max(0,j-M)}^{j}\Prob_{j-n,N-n}\Prob[S(0,i-2)=n]\nonumber\\
&=\underbrace{\sum_{j=\max(0,k-M)}^{k}\Prob_{k-j,N-j}\cdots\sum_{x=\max(0,y-M)}^{y}\Prob_{y-x,N-x}}_{i}\Prob_{x,N}.\nonumber
\end{align}

\subsection{Static Access Barring}
\label{subsec:static}

Static access barring implies that the access probability $p_i$ does not change over the burst resolution time. While it is inefficient in practice, static barring policy could serve as a baseline for the performance evaluation.

Given the QoS requirement tuple $(b^\eps, t, \eps)$, we are interested in the probability that the burst of size $N$ is still unresolved by the time $t$, and what is the backlog remaining at time $t$. So, we are looking for the bound on the backlog at the time $t$. First, we derive the MGF of the arrival and service process:
\begin{align}
&\mathbb{M}_A(\theta, 0,t) = \mathbb{E}\left[e^{\theta A(0,t)}\right] = e^{\theta N}.\label{eqn:arrival_mgf}\\
&\overline{\mathbb{M}}_S(\theta,0,t) = \E\left[e^{-\theta S(0,t)}\right]
=\sum_{k=0}^{Mt}e^{-\theta k}\Prob[S(0,t)=k]\nonumber\\&=\sum_{k=0}^{Mt}e^{-\theta k}f_{S_t}(k).\label{eqn:service_mgf}
\end{align}

\begin{figure}
	\centering
	\includegraphics[width=\linewidth]{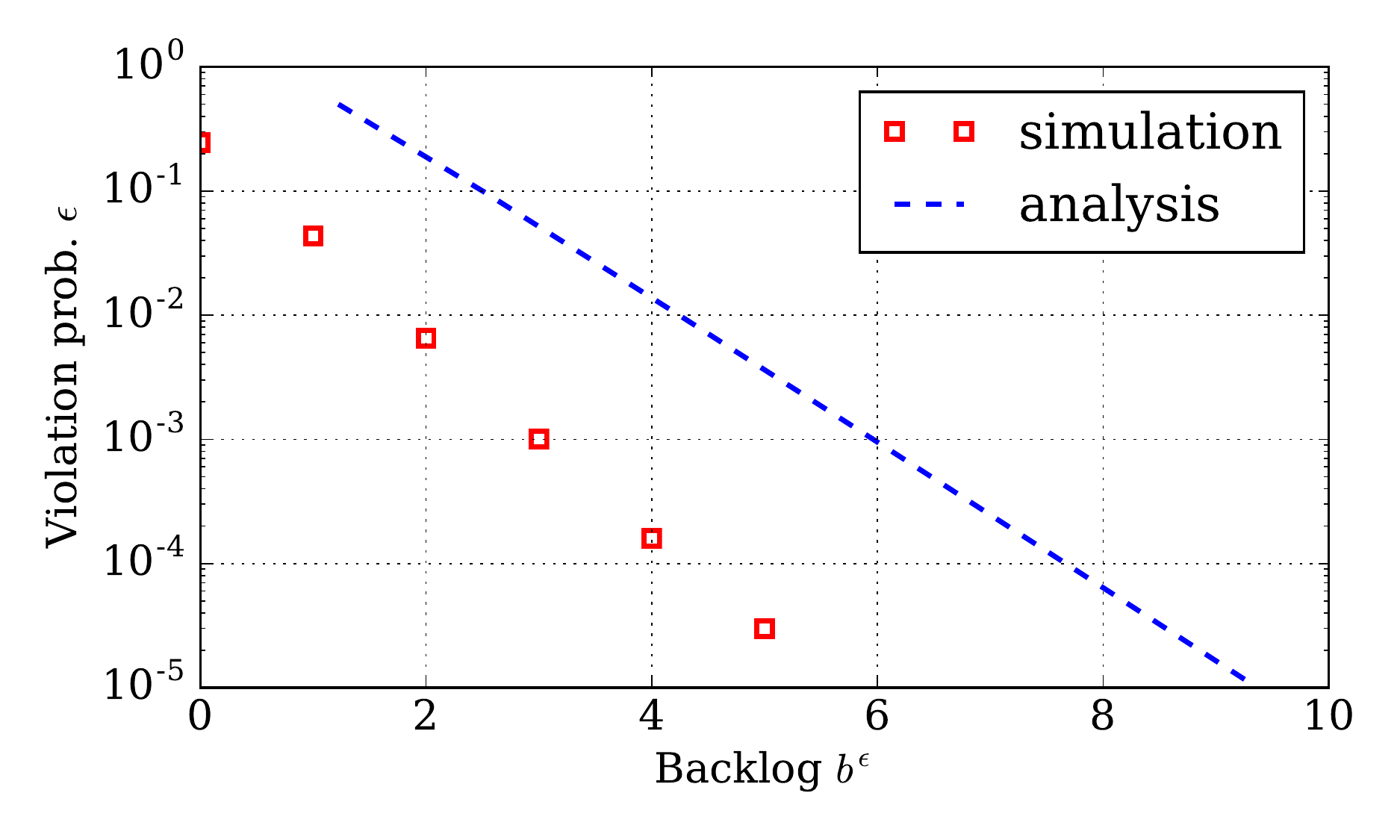}
	\caption{Static access barring: backlog $b^\eps$ vs. violation probability $\eps$. Parameters: $M=30$, $N=100$, $p=0.5$.}
	\label{fig:bounds} 
	\vspace{-0.5cm}
\end{figure}

Now, considering the time horizon $t\in[0,t)$ as a single slot, and substituting Eqns.~\eqref{eqn:arrival_mgf} and~\eqref{eqn:service_mgf} in~\eqref{eqn: backlog bound}, we can numerically compute the bound $b^\eps(t)$ for a given violation probability $\eps$ and resolution time $t$:
\begin{eqnarray}
b^\eps (t) = \inf_{\theta} \left\lbrace\frac{1}{\theta}\left(\log(e^{\theta N}\sum_{k=0}^{Mt}e^{-\theta k}f_{S_t}(k)) - \log\eps\right)\right\rbrace.\nonumber\\
\end{eqnarray}

Alternatively, we can compute the probability $\eps$ of violating a backlog bound $b^\eps$ at a given time $t$:
\begin{equation}
\eps = \inf_{\theta} \left\lbrace e^{-b^{\eps}\theta}\sum_{\tau=0}^{t}\mathbb{M}_A(\theta,\tau,t)\overline{\mathbb{M}}_S(\theta,\tau,t)\right\rbrace.
\label{eqn:epsil_by_bound}
\end{equation}
We plot a simple numerical example for the static ACB with $p=0.5$ in Fig.~\ref{fig:bounds}. In the general case of arbitrary static barring factor, computation of performance bounds via MGF calculus requires computing the cumulative service process $f_{S_t}(k)$ via Eqn.~\eqref{eqn:cumservicedist}. Hence, computing the cumulative service according to Eqn.~\eqref{eqn:cumservicedist} and then bounding the backlog distribution through network calculus has the same complexity as computing the actual backlog distribution as in~\ref{sec:prelims}. For large $N$, this becomes computationally infeasible, which motivates finding approximations for the service process.

\subsection{Dynamic Access Barring}
\label{subsec:dynamic}
From a practical point of view, dynamic access barring
presents a more interesting subject for the analysis, as it maximizes the expected efficiency of the procedure. Optimal dynamic barring policy as defined by Eqn.~\eqref{eqn:optimalacb}, is adjusting the access probability in order to maximize the expected number of successful outcomes. In other words, as the expected number of accepted UEs is a function of both access probability and the backlog, with $\E\left[b_i^\prime\right]=p_iB(i)$, dynamic access barring attempts to keep $\E[b_i^\prime]$ independent of the backlog.

This leads to an interesting observation about the expected service. The burst resolution time has now two distinct regions: first, where $B(i)\geq M$, for which is holds $p_i = \frac{M}{B(i)}$ and second, where $B(i)<M$ and $p_i=1$. In any practically relevant case, the first region is dominating the total burst resolution time, since $N\gg M$. Also, for partial burst resolution time, where the target allowed number of non-activated UEs $b^\eps\geq M$, only the first region is of interest. Here, we first derive the bound $\eps^{\text{p}}(b^\eps,t)\triangleq\eps(b^\eps, t),\forall b^\eps\geq M$ for the partial burst resolution case, and then generalize it to full burst resolution bound $\eps^{\text{f}}(0, t)\triangleq\eps(0, t)$ in the next subsection.

Given the optimal barring policy, number of devices admitted to attempt the random access in a given slot $b_i^\prime$ becomes a binomial random variable:
\begin{equation}
\Prob[b_i^\prime=x|B(i)] = \binom{B(i)}{x}{p_i^{\star}}^{x}(1-p_i^{\star})^{B(i)-x}
\label{eqn:optbarringaccess}
\end{equation}
with $\E\left[b_i^\prime\right]=p_i^{\star}B(i)=M$.

The number of admitted devices, and, hence, the service, is still dependent on the backlog, however, we can approximate the binomial distribution~\eqref{eqn:optbarringaccess} by the Poisson with the same mean.
By applying this approximation, the resulting number of admitted devices becomes independent on the backlog:
\begin{equation}
\Prob[b_i^\prime=x|B(i)] \approx \Prob[b_i^\prime=x] = \frac{M^xe^{-M}}{x!}.
\label{eqn:optbarringaccess_wc}
\end{equation}
In general, the approximation of $(n,p)$ binomial distribution with a Poisson distribution with mean $\lambda=np$ leads to an underestimation of the probability of getting a value close to the mean, and overestimating the probability of being far from the mean value. For our case, it means that we are actually underestimating the resulting service, hence, we are more conservative. For small $n$ and $p\to1$, the approximation might become even too conservative.
More importantly, using Eqn.~\eqref{eqn:optbarringaccess_wc}, and the fact that the approximated service process is independent of the current backlog state $B(i)$, we can express $\overline{\mathbb{M}}_S(\theta,\tau,t)$ via the MGF of the service in a single slot $\overline{\mathbb{M}}_S(\theta)$:
\begin{align}
&\overline{\mathbb{M}}_S(\theta,\tau,t) = \overline{\mathbb{M}}_S(\theta)^{t-\tau} =\nonumber\\ &\left(\sum_{k=0}^{M}\sum_{x=0}^{N}\frac{M^xe^{-M}}{x!}\Prob\left[s_i=k|b_i^\prime=x\right]e^{-\theta k}\right)^{t-\tau}.\label{eqn:service_final}
\end{align}
Continuing, we simplify Eqn.~\eqref{eqn:epsil_by_bound} by using~\eqref{eqn:service_final}, and obtain the violation probability bound for partial burst resolution as:
\begin{align}
&\eps^{\text{p}}(b^\eps, t) =\nonumber\\
&\inf_{\theta}\left\lbrace e^{-b^\eps\theta}\left(e^{\theta N}\overline{\mathbb{M}}_S(\theta)^{t} + \overline{\mathbb{M}}_S(\theta)\frac{1-\overline{\mathbb{M}}_S(\theta)^{t-1}}{1-\overline{\mathbb{M}}_S(\theta)}\right)\right\rbrace.\label{eqn:partial}
\end{align}
Eqns.~\eqref{eqn:service_final} and~\eqref{eqn:partial} allow us to compute the violation probability for a partial burst resolution with $b^\eps\geq M$. However, approximation~\eqref{eqn:optbarringaccess_wc} becomes very conservative as $b^\eps\to M$, hence, to provide tighter bounds we need to restrict the use of the approximation to $b^\eps > M$, and compute the remaining part of the burst resolution iteratively, which we show in the following section.

\subsection{Full Burst Resolution}
\label{sec:fullburst}
First, to control the conservativeness of the bound, we introduce a parameter $c>1$, such that we refine the split of the total burst resolution time into two regions: (1) $B(i)\geq cM$ and (2) $cM > B(i) \geq 0$.
Consider the following two random variables: $t_1$ and $t_2$, time for the backlog to be reduced from $N$ to $cM$ (region 1), and from $cM$ to $0$ (region 2), respectively. We are interested in the probability that the sum of these partial resolution times, $t_1+t_2$, is larger than the time of interest $t$.
\begin{eqnarray}
\Prob[t_1+t_2\geq t] &=&\sum_{x=0}^{t}\Prob[t_2=x]\Prob[t_1\geq t-x]\notag\\
&\leq&\sum_{x=0}^{t}\Prob[t_2=x]\eps^{\text{p}}(cM,t-x).\notag
\end{eqnarray}
By definition, $\Prob[t_1+t_2\geq t]\leq\eps^{\text{f}}(0, t)$. Hence, 
\begin{equation}
\eps^{\text{f}}(0, t)=\sum_{x=0}^{t}\Prob[t_2=x]\eps^{\text{p}}(cM,t-x),\label{eqn:fullburst}
\end{equation}
Computing the violation probability for region 1 and $t_2$ is possible using the previously introduced Eqn.~\eqref{eqn:partial}. Computing the resolution time for the region 2 can be done either using the methods for static ACB as in~\ref{subsec:static}, or directly using $\Prob_{k,n}$ as in Eqn.~\eqref{eqn:transition_prob_sending} via the recursion described in
Sec.~\ref{sec:prelims}. Since $cM\ll N$, the computational complexity is very low and proportional to $cM\times t$. Parameter $c>1$ is used to trade off conservativeness and computational complexity.

%
\section{Numerical Results}
\label{sec:evaluation}
In this section, we provide the numerical performance evaluation, and compare the analytical results with the Monte-Carlo simulation based on the Omnet++~\cite{varga2001omnet} framework. We show the results for a simultaneous activation process, where all $N$ nodes are activated at the same time $i=0$.

\begin{figure}[b!]
	\centering
	\includegraphics[width=\linewidth]{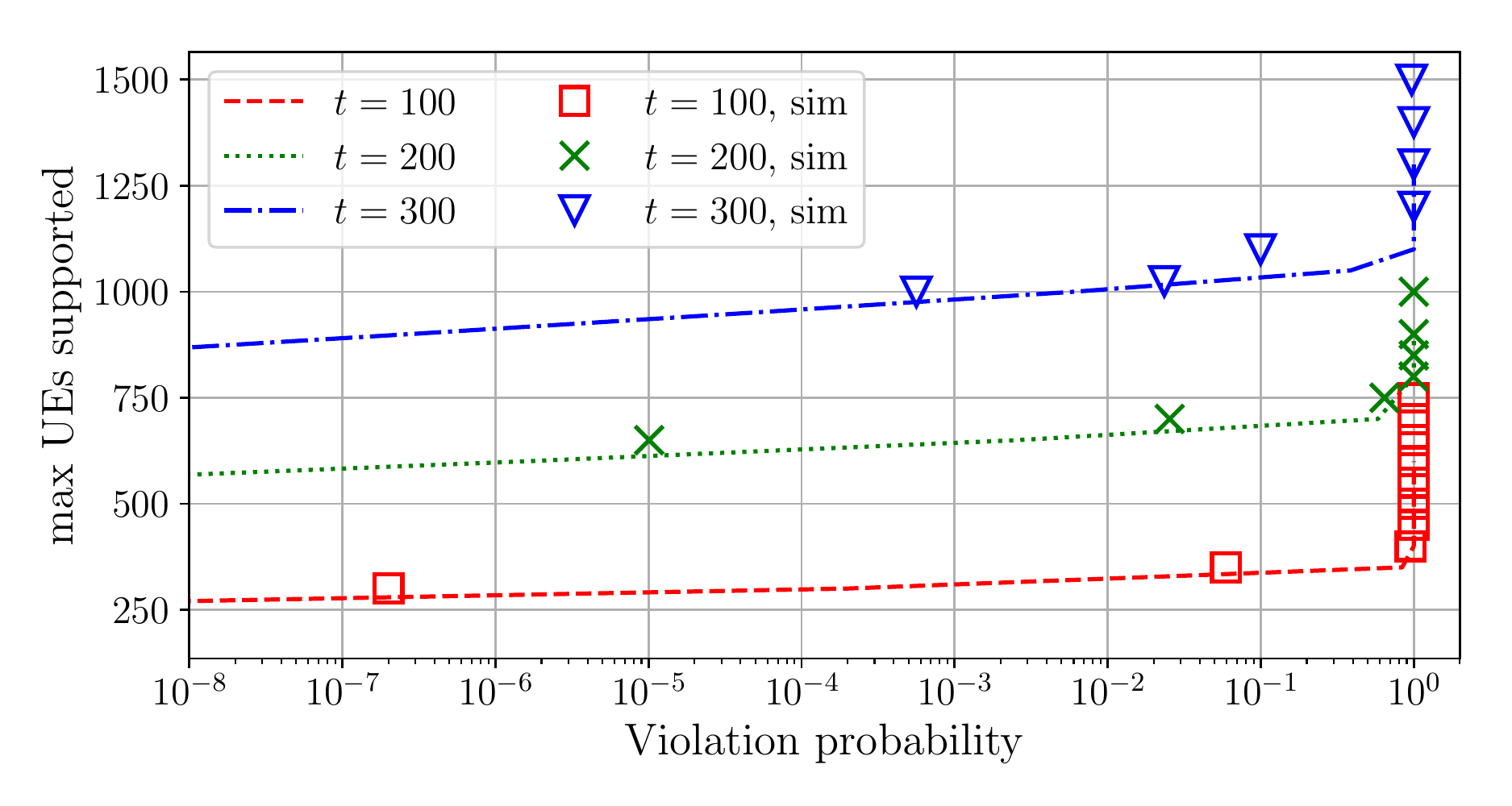}
	\caption{Maximum number of supported UEs vs. QoS requirement (total burst resolution $t$ violation probability); analysis and simulation; Parameters: $M=10$, QoS requirement $t=\{100,200,300\}$, backlog bound $b^\eps=0$, simulation with $10^7$ samples.}
	\label{fig:numerics4} 
	\vspace{-0.5cm}
\end{figure}

We first demonstrate a possible use case of the proposed model for system dimensioning. For a fixed target QoS requirement, $(b^\eps, t, \eps)$, we analytically determine the bound on the maximum number of UEs which could be supported for the requirement. The use case is illustrated in Fig.~\ref{fig:numerics4}, for the full burst resolution requirement $b^\eps$ and different resolution times $t=\{100,200,300\}$. We observe that the analytical approach provides tight lower bound for the simulations' results.

\begin{figure}[t!]
	\centering
	\includegraphics[width=\linewidth]{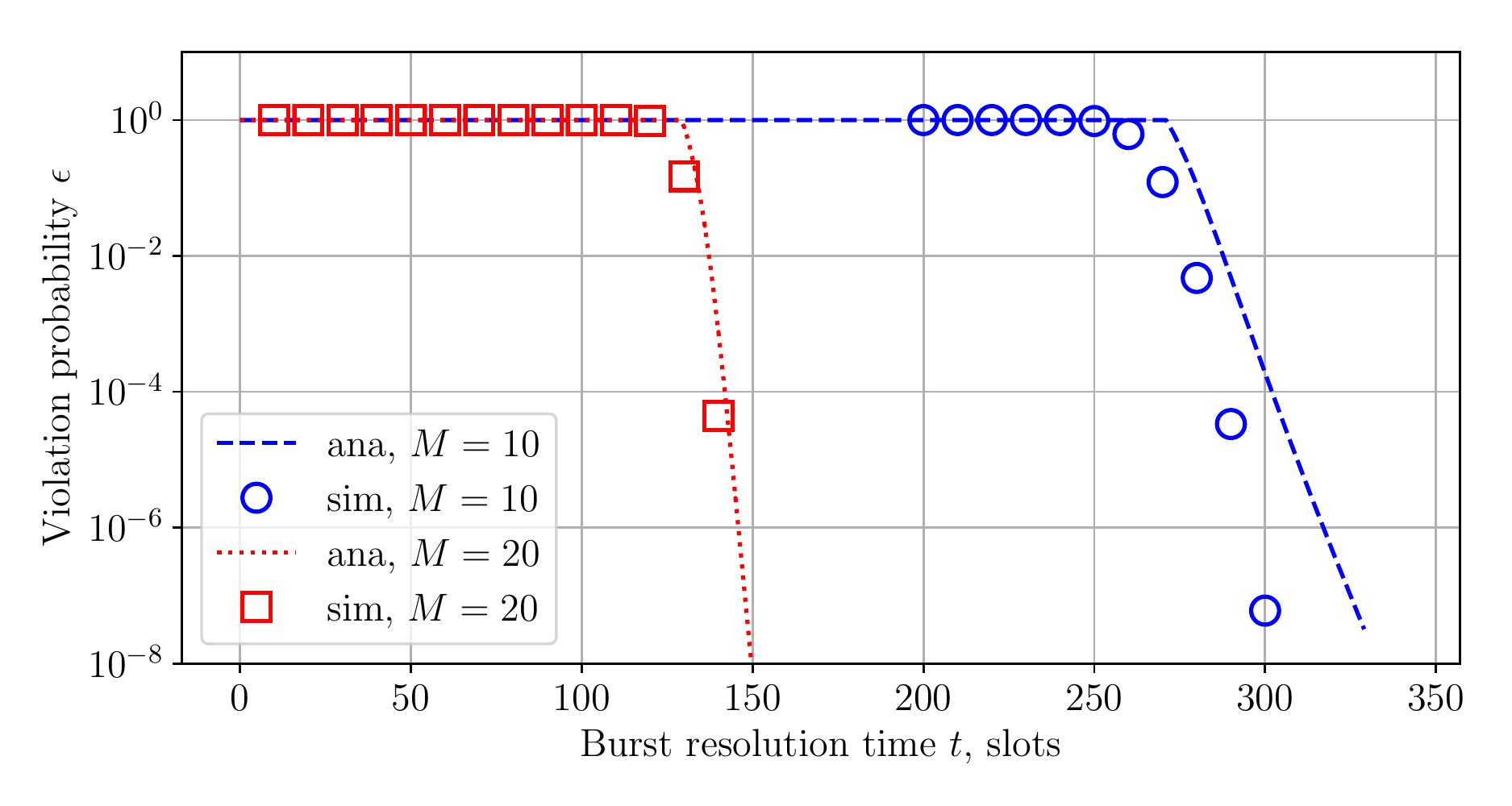}
	\caption{Partial burst resolution, backlog bound $b^\eps=3M$; minimum violation probability: analysis vs. simulation. Parameters: number of preambles $M\in\{10,20\}$, number of UEs $N=1000$. Simulations for $10^8$ samples.}
	\label{fig:numerics1} 
	\vspace{-0.5cm}
\end{figure}

\begin{figure}[t!]
	\centering
	\includegraphics[width=\linewidth]{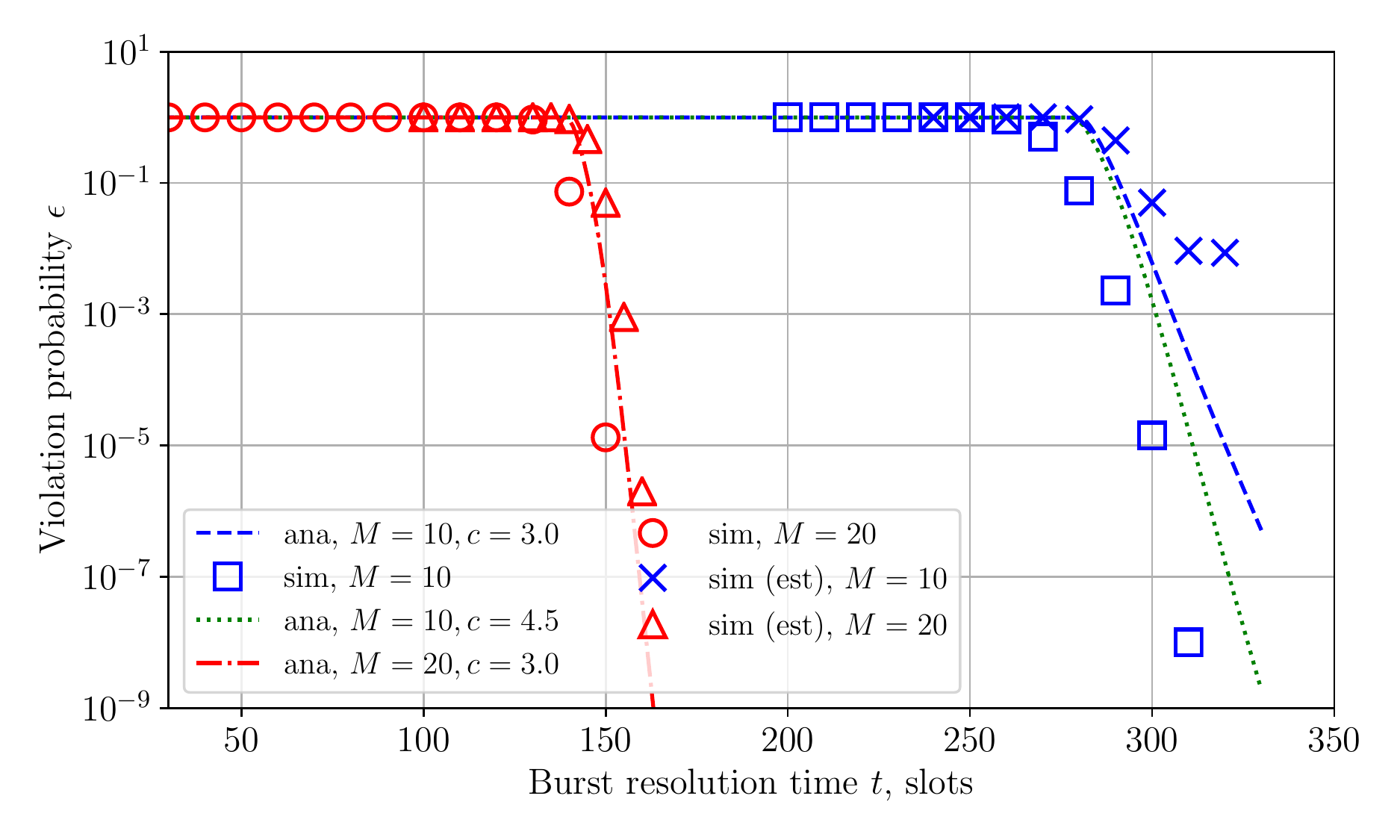}
	\caption{Full burst resolution, backlog bound $b^\eps=0$. Minimum violation probability: analysis vs. simulation. Simulations were performed with exact backlog knowledge, and with backlog estimation (denoted ``sim (est)''). Parameters: $M\in\{10,20\}$, $c\in\{3,4.5\}$, $N=1000$. Simulations for $10^8$ samples.}
	\label{fig:numerics5} 
	\vspace{-0.5cm}
\end{figure}
To further validate the analytical bounds, Fig.~\ref{fig:numerics1} depicts the violation probability for the burst resolution time $t$ for the case of partial burst resolution with the bound $b^\eps=3M$ for varying $M$. We observe that the model provides a conservative bound on violation probability for both cases, and the conservativeness increases with $t$ and decreases with $M$. For the larger number of preambles $M=20$, we observe that the slope of the CCDF is steeper than for $M=10$, indicating lower variance of the burst resolution times. 

Next, Fig.~\ref{fig:numerics5} illustrates the violation probability dependency on the resolution time for a full resolution scenario. The analytical results merge two computational models using the Eqn.~\eqref{eqn:fullburst}, at different ``splitting points'' points $cM$, with corresponding $c\in\{3.0,4.5\}$. As expected, increasing $c$ makes the overall model less conservative since the Poisson approximation of the binomial process in~\eqref{eqn:optbarringaccess_wc} becomes less conservative, at the expense of slightly longer computation.

\subsection{Impact of the Backlog Estimation} 
In some controlled scenarios, where the total number of devices $N$ and the activation pattern are known to the BS, deducing the current backlog state $B(i)$ at any time slot $i$ is possible. However, as we mentioned earlier, in many practical scenarios, backlog remains unknown to the BS, and, instead, techniques for estimating it have to be used~\cite{duan2016d,popovski2004batch,7875393,7078932,6134704}. To evaluate the impact of estimation, we relax here the assumption about the backlog state knowledge, and simulate same scenarios with the pseudo-bayesian estimation\footnote{As there is no comparison of the estimation techniques in the current state-of-the-art, we are using here the technique which has performed best in our simulated scenarios.} as proposed by Jin~\textit{et al.}~\cite{7875393}. In short, this estimation relies on the maximum-likelihood guess about the backlog $B(i)$ based on the observation of the number of idle and collided preambles in a given slot. The guess is adjusted with every new contention round.

The simulation results are also plotted in Fig.~\ref{fig:numerics5}. Comparing to the full state information case, estimation decreases the violation probability by up to almost two orders of magnitude (case $M=20$), and up to more than three orders of magnitude for the case $M=10$. The impact of estimation is higher in the second case, because the estimation relies on the observation of the number of idle preambles. When the total number of preambles is low, the estimation becomes inaccurate. Furthermore, we observe that the analytical results do not provide a bound for the case with the estimation, although they correctly capture the slope and are close to the simulation results for $M=20$. To provide an accurate performance bound for this case, future work should characterize the estimation error, and respectively include it as an offset in the definition of the serving process.
 
%
\section{Discussion and Conclusions}
\label{sec:conclusions}

In this paper, we have proposed a methodology for analyzing the reliability of the LTE random access procedure with Access Class Barring (ACB). We have considered a burst arrival scenario, where $N$ UEs are simultaneously trying to connect to the BS. For a given maximum allowed number of unconnected UEs $b^\eps$ (target backlog), and resolution time $t$, we have computed the maximum violation probability $\eps$. For dynamic access barring, we have shown that the partial burst resolution time with target backlog $b^{\eps}>M$, where $M$ is the number of available preambles, can be computed by using solely the stochastic network calculus tools. For computing full burst resolution time, we have combined iterative computation and stochastic network calculus to achieve accurate results.

The presented analysis could be useful for assessing the RA procedure performance, and integrating random access protocols into the end-to-end system reliability framework. It can also be used in standalone scenarios for system dimensioning, e.g., to decide the maximum number of UEs which could be supported for a given resolution time and reliability requirement.

Finally, as we show in Sec.~\ref{sec:evaluation}, imperfect backlog estimation has significant negative impact on the performance. This motivates further work in incorporating estimation techniques into the random access reliability analysis, as well as developing estimation techniques which can provide reliability guarantees. Additionally, future work in assessing worst-case performance of non-barring based techniques, e.g., tree algorithms~\cite{7870628,popovski2004batch,laya2014random}, and extensions of our framework for assessing other burst arrival patterns (Beta or uniform~\cite{TR37868}) are necessary.

\bibliographystyle{IEEEtran}

\begin{thebibliography}{10}
\providecommand{\url}[1]{#1}
\csname url@samestyle\endcsname
\providecommand{\newblock}{\relax}
\providecommand{\bibinfo}[2]{#2}
\providecommand{\BIBentrySTDinterwordspacing}{\spaceskip=0pt\relax}
\providecommand{\BIBentryALTinterwordstretchfactor}{4}
\providecommand{\BIBentryALTinterwordspacing}{\spaceskip=\fontdimen2\font plus
\BIBentryALTinterwordstretchfactor\fontdimen3\font minus
  \fontdimen4\font\relax}
\providecommand{\BIBforeignlanguage}[2]{{%
\expandafter\ifx\csname l@#1\endcsname\relax
\typeout{** WARNING: IEEEtran.bst: No hyphenation pattern has been}%
\typeout{** loaded for the language `#1'. Using the pattern for}%
\typeout{** the default language instead.}%
\else
\language=\csname l@#1\endcsname
\fi
#2}}
\providecommand{\BIBdecl}{\relax}
\BIBdecl

\bibitem{6815890}
A.~Osseiran, F.~Boccardi, V.~Braun, K.~Kusume, P.~Marsch, M.~Maternia,
  O.~Queseth, M.~Schellmann, H.~Schotten, H.~Taoka, H.~Tullberg, M.~A.
  Uusitalo, B.~Timus, and M.~Fallgren, ``{Scenarios for 5G mobile and wireless
  communications: the vision of the METIS project},'' \emph{IEEE Communications
  Magazine}, vol.~52, no.~5, pp. 26--35, May 2014.

\bibitem{laya2014random}
A.~Laya, L.~Alonso, and J.~Alonso-Zarate, ``{Is the Random Access Channel of
  LTE and LTE-A Suitable for M2M Communications? A Survey of Alternatives.}''
  \emph{IEEE Communications Surveys and Tutorials}, vol.~16, no.~1, pp. 4--16,
  2014.

\bibitem{TR37868}
{3GPP}, ``{Technical Report 37.868: Study on RAN Improvements for Machine-type
  Communications (Release 11)},'' {3GPP Technical Specification Group Radio
  Access Network}, Tech. Rep., 2011.

\bibitem{firstBatch1988}
I.~Cidon and M.~Sidi, ``{Conflict multiplicity estimation and batch resolution
  algorithms},'' \emph{IEEE Transactions on Information Theory}, vol.~34,
  no.~1, pp. 101--110, Jan 1988.

\bibitem{popovski2004batch}
P.~Popovski, F.~H. Fitzek, and R.~Prasad, ``Batch conflict resolution algorithm
  with progressively accurate multiplicity estimation,'' in \emph{Proceedings
  of the 2004 joint workshop on Foundations of mobile computing}.\hskip 1em
  plus 0.5em minus 0.4em\relax ACM, 2004, pp. 31--40.

\bibitem{rivest1987network}
R.~Rivest, ``Network control by bayesian broadcast,'' \emph{IEEE Transactions
  on Information Theory}, vol.~33, no.~3, pp. 323--328, 1987.

\bibitem{duan2016d}
S.~Duan, V.~Shah-Mansouri, Z.~Wang, and V.~W. Wong, ``{D-ACB: adaptive
  congestion control algorithm for bursty M2M traffic in LTE networks},''
  \emph{IEEE Transactions on Vehicular Technology}, vol.~65, no.~12, pp.
  9847--9861, 2016.

\bibitem{7875393}
H.~Jin, W.~T. Toor, B.~C. Jung, and J.~B. Seo, ``{Recursive Pseudo-Bayesian
  Access Class Barring for M2M Communications in LTE Systems},'' \emph{IEEE
  Transactions on Vehicular Technology}, vol.~66, no.~9, pp. 8595--8599, Sept
  2017.

\bibitem{wei2015modeling}
C.-H. Wei, G.~Bianchi, and R.-G. Cheng, ``{Modeling and analysis of random
  access channels with bursty arrivals in OFDMA wireless networks},''
  \emph{IEEE Transactions on Wireless Communications}, vol.~14, no.~4, pp.
  1940--1953, 2015.

\bibitem{cheng2015modeling}
R.-G. Cheng, J.~Chen, D.-W. Chen, and C.-H. Wei, ``{Modeling and analysis of an
  extended access barring algorithm for machine-type communications in LTE-A
  networks},'' \emph{IEEE Transactions on Wireless Communications}, vol.~14,
  no.~6, pp. 2956--2968, 2015.

\bibitem{7447749}
M.~Koseoglu, ``{Lower Bounds on the LTE-A Average Random Access Delay Under
  Massive M2M Arrivals},'' \emph{IEEE Transactions on Communications}, vol.~64,
  no.~5, pp. 2104--2115, May 2016.

\bibitem{7577764}
X.~Jian, Y.~Liu, Y.~Wei, X.~Zeng, and X.~Tan, ``{Random Access Delay
  Distribution of Multichannel Slotted ALOHA With Its Applications for Machine
  Type Communications},'' \emph{IEEE Internet of Things Journal}, vol.~4,
  no.~1, pp. 21--28, Feb 2017.

\bibitem{popovski2014ultra}
P.~Popovski, ``{Ultra-reliable communication in 5G wireless systems},'' in
  \emph{5G for Ubiquitous Connectivity (5GU), 2014 1st International Conference
  on}.\hskip 1em plus 0.5em minus 0.4em\relax IEEE, 2014, pp. 146--151.

\bibitem{7870628}
H.~M. G\"ursu, M.~Vilgelm, W.~Kellerer, and M.~Reisslein, ``{Hybrid Collision
  Avoidance-Tree Resolution for M2M Random Access},'' \emph{IEEE Trans. on
  Aerospace and Electronic Systems}, vol.~53, no.~4, pp. 1974--1987, Aug 2017.

\bibitem{fidler2006end}
M.~Fidler, ``An end-to-end probabilistic network calculus with moment
  generating functions,'' in \emph{Quality of Service (IWQoS). 14th IEEE
  International Workshop on}.\hskip 1em plus 0.5em minus 0.4em\relax IEEE,
  2006, pp. 261--270.

\bibitem{ciucu2014towards}
F.~Ciucu, R.~Khalili, Y.~Jiang, L.~Yang, and Y.~Cui, ``Towards a system
  theoretic approach to wireless network capacity in finite time and space,''
  in \emph{INFOCOM, 2014 Proceedings IEEE}.\hskip 1em plus 0.5em minus
  0.4em\relax IEEE, 2014, pp. 2391--2399.

\bibitem{7078932}
G.~Y. Lin, S.~R. Chang, and H.~Y. Wei, ``{Estimation and Adaptation for Bursty
  LTE Random Access},'' \emph{IEEE Transactions on Vehicular Technology},
  vol.~65, no.~4, pp. 2560--2577, April 2016.

\bibitem{6134704}
A.~Zanella, ``{Estimating Collision Set Size in Framed Slotted Aloha Wireless
  Networks and RFID Systems},'' \emph{IEEE Communications Letters}, vol.~16,
  no.~3, pp. 300--303, March 2012.

\bibitem{le2001network}
J.-Y. Le~Boudec and P.~Thiran, \emph{Network calculus: a theory of
  deterministic queuing systems for the internet}.\hskip 1em plus 0.5em minus
  0.4em\relax Springer Science \& Business Media, 2001, vol. 2050.

\bibitem{poloczek2015service}
F.~Poloczek and F.~Ciucu, ``{Service-martingales: Theory and applications to
  the delay analysis of random access protocols},'' in \emph{Computer
  Communications (INFOCOM), 2015 IEEE Conference on}.\hskip 1em plus 0.5em
  minus 0.4em\relax IEEE, 2015, pp. 945--953.

\bibitem{fidler2010survey}
M.~Fidler, ``{Survey of deterministic and stochastic service curve models in
  the network calculus},'' \emph{IEEE Communications Surveys \& Tutorials},
  vol.~12, no.~1, pp. 59--86, First 2010.

\bibitem{al2013min}
H.~Al-Zubaidy, J.~Liebeherr, and A.~Burchard, ``A (min,$\times$) network
  calculus for multi-hop fading channels,'' in \emph{INFOCOM, 2013 Proceedings
  IEEE}.\hskip 1em plus 0.5em minus 0.4em\relax IEEE, 2013, pp. 1833--1841.

\bibitem{varga2001omnet}
A.~Varga \emph{et~al.}, ``{The OMNeT++ discrete event simulation system},'' in
  \emph{Proceedings of the European simulation multiconference (ESM’2001)},
  vol.~9, no. S 185.\hskip 1em plus 0.5em minus 0.4em\relax sn, 2001, p.~65.

\end{thebibliography}

\end{document}